# Fault Masking By Probabilistic Voting

B. Baykant ALAGÖZ

**Abstract:** *In this study, we introduced a probabilistic voter, regarding symbol probabilities in decision process besides majority consensus. Conventional majority voter is independent of functionality of redundant modules. In our study, proposed probabilistic voter is designed corresponding to functionality of the redundant module. We tested probabilistic voter for 3 and 5 redundant modules with random transient errors inserted the wires and it was seen from simulation results that Multi-Modular Redundancy (M-MR) with Probabilistic Voting (PV) had been shown better availability performance than conventional majority voter.*

**Keyword:** Fault Tolerance, Fault Masking, Voting Algorithm.

## Introduction:

Majority voting is the most common voting technique used in Multi-Modular Redundancy (M-MR) for fault masking in digital systems. [1-4] The most basic one is bit-by-bit majority voting, in which voter algorithm selects the most common output. In the Table 1, truth table of conventional bit-by-bit majority voting for Triple Modular Redundancy (TMR) was given to demonstrate output selection from redundant modules according to majority consensus. In this table, $y_1$, $y_2$ and $y_3$ are output of modules and $y$ represents voter output. General architecture of M-MR system was shown in the Figure 1. In older studies, conventional majority voter was seen to capable of masking single errors which occurs when one of three modules is faulty.

**Table 1.** Conventional bit-by-bit majority voting [3]

| $y_1$ | $y_2$ | $y_3$ | $y$ |
|---|---|---|---|
| 0 | 0 | 0 | 0 |
| 0 | 0 | 1 | 0 |
| 0 | 1 | 0 | 0 |
| 0 | 1 | 1 | 1 |
| 1 | 0 | 0 | 0 |
| 1 | 0 | 1 | 1 |
| 1 | 1 | 0 | 1 |
| 1 | 1 | 1 | 1 |

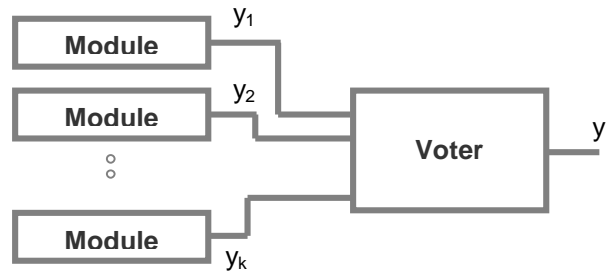

**Figure 1.** Block diagram of the Multi-Modular Redundancy

Conventional bit-by-bit majority voting made its decision according to majority consensus at outputs of redundant modules to select the most reliable results. It does not consider any information based on symbol probabilities derived from functionality of modules. In this paper, we proposed a bit-by-bit voting algorithm, which is rather aware of module functionality, to increase availability of M-MR. This awareness relies on error probability of obtained result at output of the modules. In this manner, we called proposed voting algorithm as probabilistic voting. Probabilistic voting considerably raised availability of M-MR in our simulations and also reduced the complexity of the voter circuitry.

In the next sections, we will suggest a probabilistic voter design methodology and then we will do performance comparison with conventional majority voter designed for TMR, made of three redundant modules, and also 5MR, made of five redundant modules.

**Probabilistic Voting:**

Let member of set $\{0,1\}$ be output symbols of redundant modules $y_1, y_2, y_3...y_k$. $E_0$ is probability of symbol $\{0\}$ being an error at output and $E_1$ is probability of symbol $\{1\}$ being an error at output of module. For redundant module with $n$-bit input, logic function of modules is defined by $2^n$ term in its truth table. Number of terms resulting $\{0\}$ at output of module is denoted by $N_0$. If symbol $\{1\}$ at output is an error, probability of this error can be expressed by

$$E_1 = \frac{N_0}{2^n} \quad (1)$$

Number of terms resulting $\{1\}$ at output of module is denoted by $N_1$ and similarly, if symbol $\{0\}$ at output is an error, probability of this error can be expressed by,

$$E_0 = \frac{N_1}{2^n} \quad (2)$$

Lets expand truth table of bit-by-bit voter seen in Table 1 to Table 2.

**Table 2.** Generic truth table of probabilistic voting for TMR

| $y_1$ | $y_2$ | $y_3$ | Cost for 0 $C_0$ | Cost for 1 $C_1$ | $y$ |
|---|---|---|---|---|---|
| 0 | 0 | 0 | $E_0/3$ | $\infty$ | 0 |
| 0 | 0 | 1 | $E_0/2$ | $E_1$ | X |
| 0 | 1 | 0 | $E_0/2$ | $E_1$ | X |
| 0 | 1 | 1 | $E_0$ | $E_1/2$ | X |
| 1 | 0 | 0 | $E_0/2$ | $E_1$ | X |
| 1 | 0 | 1 | $E_0$ | $E_1/2$ | X |
| 1 | 1 | 0 | $E_0$ | $E_1/2$ | X |
| 1 | 1 | 1 | $\infty$ | $E_1/3$ | 1 |

For a given particular logic function, '$X$' values in generic truth tables are determined by following equation.

$$X = \begin{cases} 1 & C_1 \leq C_0 \\ 0 & C_0 < C_1 \end{cases} \quad (3)$$

Here, $C_o$ is cost function for symbol $\{0\}$ and $C_1$ is cost function for symbol $\{1\}$. They can be calculated by following formulas,

$$C_0 = \frac{E_0}{V_0} \text{ and } C_1 = \frac{E_1}{V_1} \qquad (4)$$

Here $V_0$ is number of redundant modules yielding $\{0\}$ symbol at output and $V_1$ is number of redundant modules yielding $\{1\}$ symbol at output. For the TMR, $(V_0, V_1)$ couples can take values given by the set of $\{(0,3),(1,2),(2,1),(3,0)\}$.

Probabilistic voting selects the output symbol, which has lower cost by mean of equation (3). Cost of each symbol is calculated depending onto probability of obtained symbol being an error ($E_0, E_1$) and number of modules producing this symbol ($V_0, V_1$) as expressed at equation (4). Probability of obtained symbol being an error was expressed by equation (1) and equation (2) for each symbol under assumption of the uniform distributed random inputs applied to modules and under this assumption, ($E_0, E_1$) parameters is very dependent on the functionality of the redundant module. On the other hand, number of modules yielding a symbol reflects strength of symbol among redundant modules $y_1, y_2, y_3...y_k$ and cost function falls by strength of symbol. Since, lower cost means lower error probability and higher consensus, probabilistic voter selects the symbol that has lower cost as the most reliable output symbol.

## Probabilistic Voter Design Examples:

In this section, we will demonstrate a design example of probabilistic voters of TMR for a given logic function. We can use Table 2 that is a generic truth table, subjecting to $E_1$ and $E_0$. Lets consider 4-input logic function, of which truth table was given in Table 3.

**Table 3.** Truth table of $f$ function

| $a$ | $b$ | $c$ | $d$ | $f$ |
|---|---|---|---|---|
| 0 | 0 | 0 | 0 | 0 |
| 0 | 0 | 0 | 1 | 0 |
| 0 | 0 | 1 | 0 | 0 |
| 0 | 0 | 1 | 1 | 0 |
| 0 | 1 | 0 | 0 | 0 |
| 0 | 1 | 0 | 1 | 0 |
| 0 | 1 | 1 | 0 | 0 |
| 0 | 1 | 1 | 1 | 0 |
| 1 | 0 | 0 | 0 | 0 |
| 1 | 0 | 0 | 1 | 0 |
| 1 | 0 | 1 | 0 | 0 |
| 1 | 0 | 1 | 1 | 0 |
| 1 | 1 | 0 | 0 | 1 |
| 1 | 1 | 0 | 1 | 0 |
| 1 | 1 | 1 | 0 | 1 |
| 1 | 1 | 1 | 1 | 0 |

Number of symbol {0} in the table is 14.($N_0 = 14$) Number of symbol {1} in the table is 2.($N_1 = 2$) Number of input bits is 4. ($n = 4$). Using these parameters coming from logic function, one can calculates error probability for symbol {1} as $E_1 = 14/16$ and for symbol {0} as $E_0 = 2/16$. Now, we can construct truth table of probabilistic voter for the logic function given in Table 3. Considering Table 2, produced truth table of probabilistic voter was given in the Table 4.

**Table 4.** Truth table for Probabilistic voting

| $y_1$ | $y_2$ | $y_3$ | Cost for {0} ($C_0$) | Cost for {1} ($C_1$) | $y$ |
|---|---|---|---|---|---|
| 0 | 0 | 0 | *2/48* | ∞ | 0 |
| 0 | 0 | 1 | *2/32* | *14/16* | 0 |
| 0 | 1 | 0 | *2/32* | *14/16* | 0 |
| 0 | 1 | 1 | *2/16* | *14/32* | 0 |
| 1 | 0 | 0 | *2/32* | *14/16* | 0 |
| 1 | 0 | 1 | *2/16* | *14/32* | 0 |
| 1 | 1 | 0 | *2/16* | *14/32* | 0 |
| 1 | 1 | 1 | ∞ | *14/48* | 1 |

Logic equation of probabilistic voter originated from Table 4 can be written as following,

$$y = y_1 \cdot y_2 \cdot y_2 \qquad (5)$$

## Simulation Results For Probabilistic Voter:

Faults on logic gates were assumed to result an error bit at the output of the gate. This error bit would be expectedly complement of correct result. In simulations, error bits satisfying desired error probabilities were inserted to wires, which is connected to outputs of faulty gates. Program for this error insertion mechanism used in the simulation is seen below.

**Program:** Error insertion to wire

```
% rand is random number between 0.0000 and 1.0000
% Pe: error probability for wire. net: Variable for wire

  if (rand < Pe)
      net = not(net);
  end if
```

In the simulation, 5000 inputs with uniform distribution were applied to faulty redundant modules for each error probability given in a range. Outputs of faulty redundant modules were connected both of conventional majority voter and probabilistic voter. Correct results were counted for availability calculation defined by following equation,

$$A = \frac{Cr}{Tr} \qquad (6)$$

$Cr$ is number of correct output and $Tr$ is number of total output. $A$ is the availability of the system.

In the Figure 2, availability of the faulty module, availability of TMR with conventional majority voter and availability of TMR with probabilistic voter are compared for error probability of wires ranging 0.001 to 0.5.

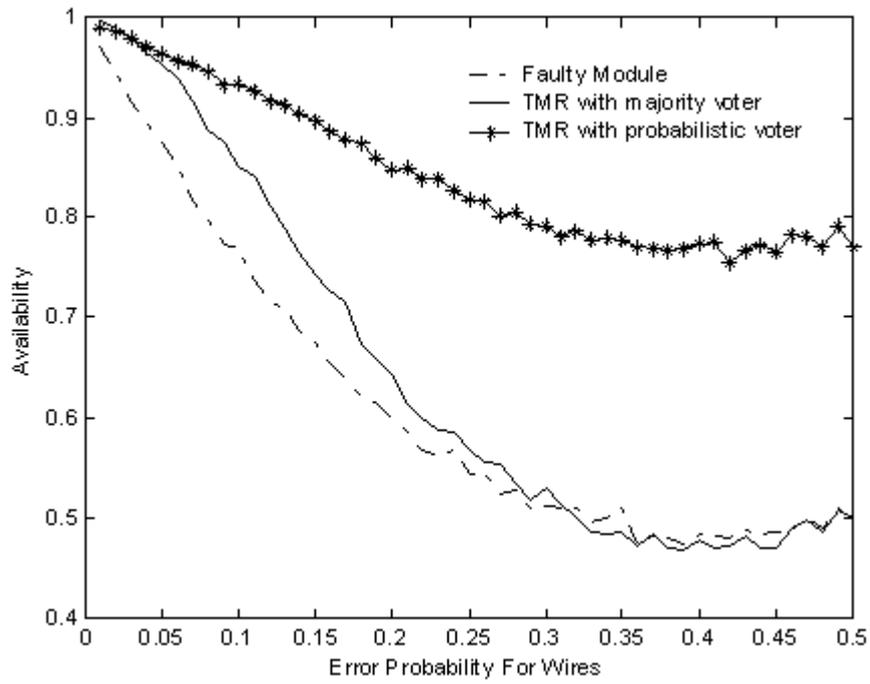

**Figure 2.** Availability comparison

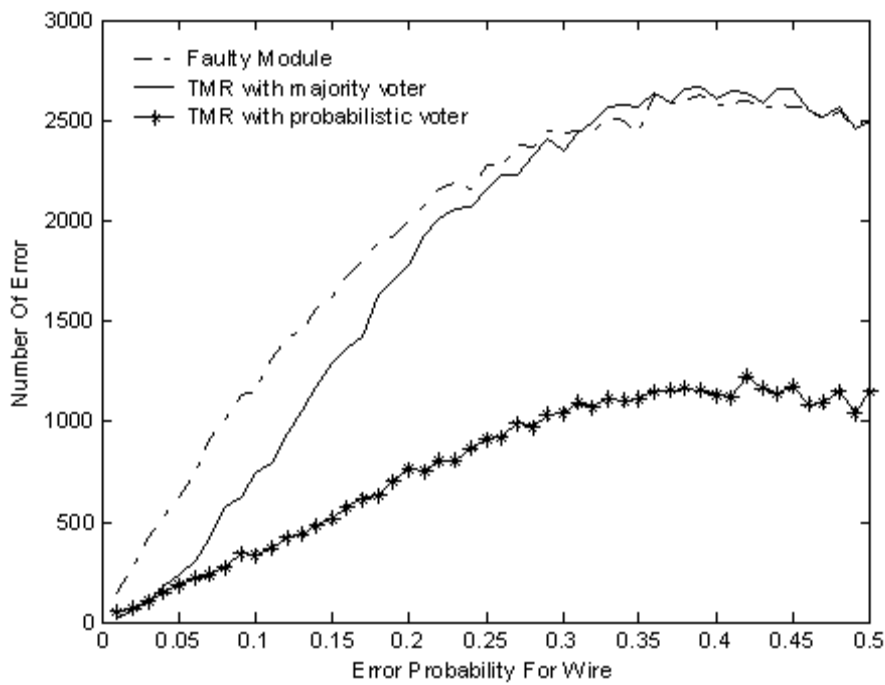

**Figure 3.** Number of errors at output of system

It is seen from Figure 2 and Figure 3, probabilistic voter exhibited superior fault masking performance than conventional bit-by-bit majority voter. Besides, voter circuitry for probabilistic voter has lower complexity than conventional majority voter.

## Probabilistic Voter Design For 5MR:

Generic truth table of probabilistic voting for 5MR was given in Table 5.

**Table 5.** Generic truth table of probabilistic voting for 5MR

| $y_1$ | $y_2$ | $y_3$ | $y_4$ | $y_5$ | $C_0$ | $C_1$ | $y$ |
|---|---|---|---|---|---|---|---|
| 0 | 0 | 0 | 0 | 0 | $E_0/5$ | $\infty$ | 0 |
| 0 | 0 | 0 | 0 | 1 | $E_0/4$ | $E_1$ | X |
| 0 | 0 | 0 | 1 | 0 | $E_0/4$ | $E_1$ | X |
| 0 | 0 | 0 | 1 | 1 | $E_0/3$ | $E_1/2$ | X |
| 0 | 0 | 1 | 0 | 0 | $E_0/4$ | $E_1$ | X |
| 0 | 0 | 1 | 0 | 1 | $E_0/3$ | $E_1/2$ | X |
| 0 | 0 | 1 | 1 | 0 | $E_0/3$ | $E_1/2$ | X |
| 0 | 0 | 1 | 1 | 1 | $E_0/2$ | $E_1/3$ | X |
| 0 | 1 | 0 | 0 | 0 | $E_0/4$ | $E_1$ | X |
| 0 | 1 | 0 | 0 | 1 | $E_0/3$ | $E_1/2$ | X |
| 0 | 1 | 0 | 1 | 0 | $E_0/3$ | $E_1/2$ | X |
| 0 | 1 | 0 | 1 | 1 | $E_0/2$ | $E_1/3$ | X |
| 0 | 1 | 1 | 0 | 0 | $E_0/3$ | $E_1/2$ | X |
| 0 | 1 | 1 | 0 | 1 | $E_0/2$ | $E_1/3$ | X |
| 0 | 1 | 1 | 1 | 0 | $E_0/2$ | $E_1/3$ | X |
| 0 | 1 | 1 | 1 | 1 | $E_0$ | $E_1/4$ | X |
| 1 | 0 | 0 | 0 | 0 | $E_0/4$ | $E_1$ | X |
| 1 | 0 | 0 | 0 | 1 | $E_0/3$ | $E_1/2$ | X |
| 1 | 0 | 0 | 1 | 0 | $E_0/3$ | $E_1/2$ | X |
| 1 | 0 | 0 | 1 | 1 | $E_0/2$ | $E_1/3$ | X |
| 1 | 0 | 1 | 0 | 0 | $E_0/3$ | $E_1/2$ | X |
| 1 | 0 | 1 | 0 | 1 | $E_0/2$ | $E_1/3$ | X |
| 1 | 0 | 1 | 1 | 0 | $E_0/2$ | $E_1/3$ | X |
| 1 | 0 | 1 | 1 | 1 | $E_0$ | $E_1/4$ | X |
| 1 | 1 | 0 | 0 | 0 | $E_0/3$ | $E_1/2$ | X |
| 1 | 1 | 0 | 0 | 1 | $E_0/2$ | $E_1/3$ | X |
| 1 | 1 | 0 | 1 | 0 | $E_0/2$ | $E_1/3$ | X |
| 1 | 1 | 0 | 1 | 1 | $E_0$ | $E_1/4$ | X |
| 1 | 1 | 1 | 0 | 0 | $E_0/2$ | $E_1/3$ | X |
| 1 | 1 | 1 | 0 | 1 | $E_0$ | $E_1/4$ | X |
| 1 | 1 | 1 | 1 | 0 | $E_0$ | $E_1/4$ | X |
| 1 | 1 | 1 | 1 | 1 | $\infty$ | $E_1/5$ | 1 |

Let design probabilistic voter for 5MR with redundant module given as,

$$f = a \cdot b \cdot c \cdot \overline{d} + a \cdot b \cdot \overline{c} \cdot \overline{d} + \overline{a} \cdot \overline{b} \cdot c \cdot \overline{d} + \overline{a} \cdot \overline{b} \cdot c \cdot d \qquad (7)$$

Relevant parameters for $f$ function would be $E_1 = 12/16$ and $E_0 = 4/16$, logic equation of probabilistic voter inferred from Table 5 can be written as following,

$$y = \bar{y}_1 \cdot y_2 \cdot y_3 \cdot y_4 \cdot y_5 + y_1 \cdot \bar{y}_2 \cdot y_3 \cdot y_4 \cdot y_5 + y_1 \cdot y_2 \cdot \bar{y}_3 \cdot y_4 \cdot y_5$$
$$+ y_1 \cdot y_2 \cdot y_3 \cdot \bar{y}_4 \cdot y_5 + y_1 \cdot y_2 \cdot y_3 \cdot y_4 \cdot \bar{y}_5 + y_1 \cdot y_2 \cdot y_3 \cdot y_4 \cdot y_5$$

(8)

Obtained results from the simulation are illustrated in the Figure 4 and Figure 5.

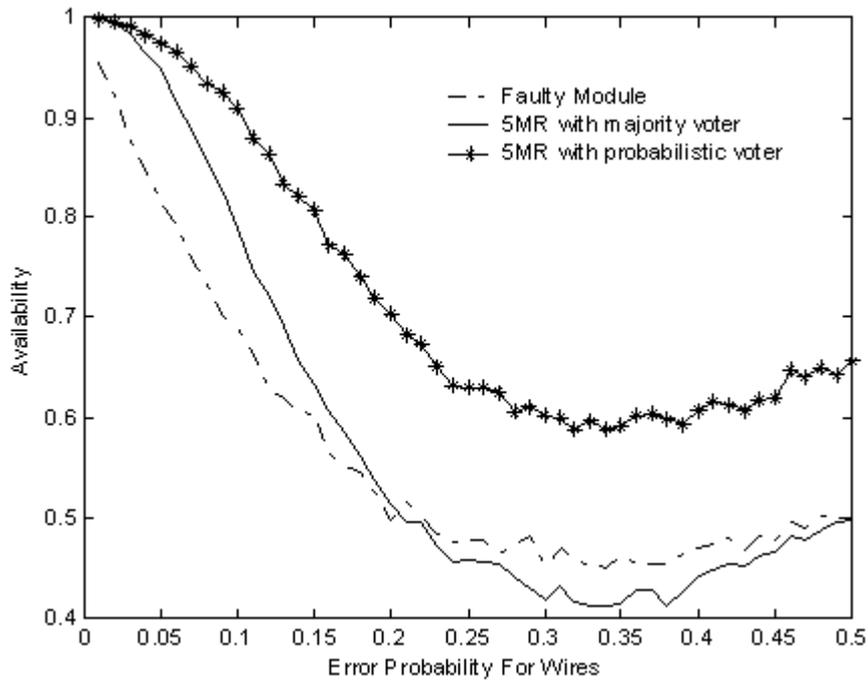

**Figure 4.** Availability comparison

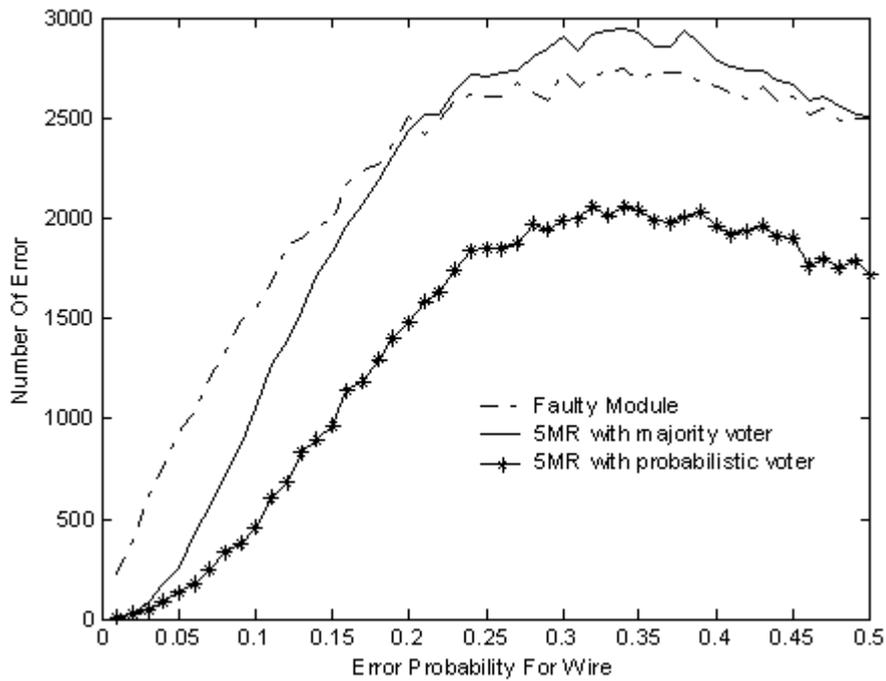

**Figure 5.** Number of errors at output of system

## Conclusions:

Probabilistic voting regarding symbol probabilities as well as majority consensus was seen to provide better availability and error masking performance.

In the paper, we introduced a probabilistic voter design approach for digital systems. Suggested probabilistic voter were demonstrated to able to have better performance in case of uniform distributed random error insertion to wires and under application of uniformly distributed random inputs. Other important point to be consider that, complexity of the probabilistic voter can be lower than conventional majority voter circuitry. This point gains importance when high volume logic is needed to support by M-MR particularly in Register Transfer Level (RTL) designs. As a consequence, probabilistic voter designed for a specific logic function can have better availability-complexity performance than conventional majority voters designed for common-use.

*…to memory of my brother Serdar Onur Alagöz.*